\def\be{\begin{equation}}
\def\ee{\end{equation}}
\def\ba{\begin{eqnarray}}
\def\ea{\end{eqnarray}}
\def\no{\nonumber \\}
\newcommand{\nid}{\noindent}

\documentclass{PoS}

\title{Mass Splitting of Staggered Fermion 
and $SO(2D)$ Clifford Algebra}

\ShortTitle{Mass Splitting of Staggered Fermion 
and $SO(2D)$ Clifford Algebra}

\author{\speaker{Morio Hatakeyama}, Hideyuki Sawanaka\\ 
Graduate School of Science and Technology, Niigata University\\ 
Ikarashi 2-8050, Niigata 950-2181, Japan\\
E-mail: \email{hatake@muse.sc.niigata-u.ac.jp}, 
\email{hide@muse.sc.niigata-u.ac.jp}}
\author{Hiroto So\thanks{This work is supported in part by 
the Grants-in-Aid for Scientific Research No.  17540242 from 
the Japan Society for the Promotion of Science.}\\
Department of Physics, 
Niigata University\\
Ikarashi 2-8050, Niigata 950-2181, Japan\\
E-mail: \email{so@muse.sc.niigata-u.ac.jp}}

\abstract{We present a new method to introduce rotationally 
invariant terms in staggered fermions 
which is based on an $SO(2D)$ Clifford algebra formulation,
where $D$ means the number of space-time dimensions.  
We have four candidates for improved mass terms that can split the degenerate mass of staggered fermions.
Among them, we analyze three types of combinations and find only one case
that can identify with the light single Dirac mode.}

\FullConference{XXIV International Symposium on Lattice Field Theory\\
		 July 23-28 2006\\
		 Tucson Arizona, USA}

\usepackage{bm}
\begin{document}
\section{Introduction}    
Staggered fermions are formulated 
in which  species doublers of a Dirac field 
are interpreted as physical degrees of freedom, {\it tastes}, on 
lattice~\cite{KS,Susskind}.   
However, it remains for a 4-fold degeneracy problem  of tastes 
in four dimensions to be unsolved.  
Although a fourth-root trick of the determinant 
in a staggered Dirac operator  is  an approach to unfold the degeneracy 
and studies on  its theoretical basis 
are developed~\cite{Adams:2003rm,Shamir,BGS}, 
we have no local expression of 
one taste Dirac fermion after the fourth-root trick.  

Avoiding the trick, there are pioneering works 
for solving the degeneracy tried 
by improved staggered fermion approaches~\cite{Mitra,Hart}.  
The improved actions generally include more operators than 
the original staggered one and are difficult to treat them~\cite{Toolkit}.  
For the control of their operators, 
we make use of staggered fermions 
on a $D$-dimensional lattice space based on 
an $SO(2D)$ Clifford algebra, 
and a discrete rotational symmetry can be represented by the algebra~\cite{IKMSS}.

In this article, 
to split degenerate tastes, 
we add new four operators to the original staggered action in two dimensions.  
Only these four operators keep the discrete rotational symmetry 
in any dimension~\cite{IKMSS}.  
The total mass matrix analysis 
is insufficient because  the matrix  does not commute with  
the kinetic term.  Therefore, we also analyze 
the propagator and the pole of the improved free staggered Dirac operator.  
It is found that only one combination 
in these operators is a good candidate after these analyses.  
More details can be found in Ref.~\cite{HSS}.  

\section{Formulation of Staggered Fermions and  Rotational Symmetry} 
The formulation of staggered fermions 
on the $D$-dimensional lattice space has been presented 
based on the $SO(2D)$ Clifford algebra~\cite{IKMSS}.  The basic idea 
is that the dimension of the total representation space 
including  spinor and taste spaces, $2^D$ 
is the same as that of  an $SO(2D)$ spinor representation.  $2^D$ 
is also the same as the number of sites in a $D$-dimensional 
hypercube.  To avoid the double counting of sites, 
the lattice coordinate $n_{\mu}$  is  noted by 
\be
 n_{\mu} = 2 N_{\mu}+ c_{\mu} + r_{\mu},  
\ee
\nid
where $N_{\mu}$ is the global coordinate of the hypercube.  
In this case, a fundamental unit is $2a$, where $a$ is a lattice 
constant, and is set to unity.  $ c_{\mu} = 1/2$ for any $\mu$ 
means  the coordinate of a center in the $D$-dimensional hypercube and 
$r_\mu$ does the relative coordinate of a site to the center.  
The relative coordinate is the same as  a weight of  the spinor 
representation in  $SO(2D)$.  

Although our formulation can be generalized,  we consider a free theory 
in a two-dimensional lattice, for simplicity.  Relative 
coordinates of four sites around a plaquette are written by 
\be
 (r_1, r_2) = (-1/2, -1/2),~ (-1/2, 1/2),~ (1/2, -1/2),~ (1/2, 1/2) .
\label{4sites}
\ee
\nid
Actually, our staggered fermion is defined on sites (\ref{4sites}) as 
\be
 \Psi(n) \equiv \Psi_{r}(N) = \left(\begin{array}{c} 
  \Psi_{(-1/2,-1/2)} \\ \Psi_{(-1/2,1/2)} \\ 
  \Psi_{(1/2,-1/2)} \\ \Psi_{(1/2,1/2)} \end{array}\right) (N) 
 \equiv  \left(\begin{array}{c} 
  \Psi_1 \\ \Psi_2 \\ \Psi_3 \\ \Psi_4 \end{array}\right) (N). 
\ee
\nid
It is noted that $\Psi_1$ and  $\Psi_4$ are put on even sites and 
$\Psi_2$ and  $\Psi_3$ are put on odd sites.  

An $SO(4)$ Clifford algebra plays a crucial role in 
two-dimensional cubic lattice formulations~\cite{IKMSS}.  The original 
staggered fermion action~\cite{KS,Susskind} can be written as 
\be
 S_{st} = \sum_{N,N^\prime,r,r^\prime,\mu,\vec{\tau}} 
 \bar{\Psi}_r(N)(D_{\mu}^{\vec{\tau}})_{(N,N^\prime)} 
 (\Gamma_{\mu,\vec{\tau}})_{(r,r^\prime)}\Psi_{r^\prime}(N^\prime) , 
\label{action}
\ee
\nid
where $\vec{\tau}$ is a two-dimensional vector 
with its components of $\pm 1/2$  and 
$ D_{\mu}^{\vec{\tau}}$ for $\mu = 1, 2$, is 
a generalized difference  operator defined by 
\be
 (D_{\mu}^{\vec{\tau}})_{(N,N^\prime)} \equiv 
 \frac{1}{2^2}\sum_{\vec{\sigma}=0,1} 
 (-1)^{(\vec{c}+\vec{\tau})\cdot\vec{\sigma}} 
 (\nabla_\mu^{\vec{\sigma}})_{(N,N^\prime)} , 
\ee
\nid
with
\be
 (\nabla_\mu^{\vec{\sigma}})_{(N,N^\prime)} = \left\{\begin{array}{ll} 
  \delta_{N,N^\prime} 
- 
  \delta_{N-\hat{\mu},N^\prime} 
  \equiv \nabla_\mu^-, & ~\sigma_\mu=0, \\
 \\ 
  \delta_{N+\hat{\mu},N^\prime} 
- 
  \delta_{N,N^\prime}
  \equiv\nabla_\mu^+, & ~\sigma_\mu=1. 
 \end{array}\right. 
\label{gen-diff}
\ee
\nid
$\vec{\sigma}$ is a two-dimensional vector dual to $\vec{\tau}$ and 
$\nabla_\mu^+$ ($\nabla_\mu^-$) implies 
a forward (backward) difference operator along the $\mu$-direction, 
respectively.  
The matrix 
$\Gamma_{\mu,\vec{\tau}}$ in our action (\ref{action}) is 
composed of the $SO(4)$ Clifford algebra 
$\Gamma_{\mu,-\vec{c}} \equiv \gamma_{\mu}$ and 
$\Gamma_{\mu,-\vec{c}+\vec{e}_\mu} \equiv i\tilde{\gamma}_{\mu}$, 
\be
(\Gamma_{\mu,\vec{\tau}})_{(r,r^\prime)} \equiv 
\left\{\begin{array}{ll} 
((\sigma_3^{1/2+\tau_1} \otimes \sigma_3^{1/2+\tau_2}) 
\times 
(\sigma_1 \otimes \bm{1}))_{(r,r^\prime)}~, ~~\mu=1, \\ 
 \\
((\sigma_3^{1/2+\tau_1} \otimes \sigma_3^{1/2+\tau_2}) 
\times 
(\sigma_3 \otimes \sigma_1))_{(r,r^\prime)}~, ~~\mu=2, 
 \end{array}\right.
\ee
\nid
where $\vec{e}_\mu$ is the unit vector along the $\mu$-direction.  
Here we denote the fundamental algebra, or the $SO(4)$ Clifford algebra as 
\be
 \{\gamma_{\mu},\gamma_{\nu}\} = 
 \{\tilde{\gamma}_{\mu},\tilde{\gamma}_{\nu}\} = 
2\delta_{\mu\nu}~,~~
 \{\gamma_{\mu},\tilde{\gamma}_{\nu}\}=0.  
\ee
For a discrete rotation with angle $\pi/2$  around the center, 
the transformations of global and relative coordinates are denoted by 
$N \rightarrow  R(N)$,~~$r \rightarrow  R(r)$,~~and that of fermion  is 
\be
 \Psi(N) \rightarrow V_{12} \Psi(R(N)).
\ee
\nid
$V_{12}$ is  a rotation matrix about a spinor index 
in the $SO(4)$ base, up to a phase factor given by a form 
\be
 V_{12} = \frac{e^{i\vartheta}}{2} \Gamma_5 
 (\tilde{\gamma}_1-\tilde{\gamma}_2)(1+\gamma_1\gamma_2), 
\ee
\nid
where 
$\Gamma_5 \equiv \gamma_1\gamma_2\tilde{\gamma}_1\tilde{\gamma}_2 
= diag(1,-1,-1,1)$.  
Only the following four operators $\bar{\Psi} O_i \Psi$ for $i=1, 2, 3, 4$, 
\be
 O_1 = \bm{1}, \quad 
 O_2 = i\gamma_1\gamma_2 \equiv \Gamma_3, \quad 
 O_3 = \tilde{\gamma}_1 +  \tilde{\gamma}_2, \quad 
 O_4 = \Gamma_3 (\tilde{\gamma}_1 + \tilde{\gamma}_2), 
\label{ops}
\ee
\nid
are invariant under the rotation $V_{12}O_i V_{12}^\dag$.  
Our analyses in the following sections concentrate on 
the improved staggered fermion action  by these four matrices.  

\section{Analysis of Mass Matrices} 
To split masses in desired degenerate tastes  we introduce four
rotationally invariant operators  which we denote as 
$\bar{\Psi} O_i \Psi$~\cite{IKMSS}, 
for the original staggered fermion action~(\ref{action}).  
The total mass matrix form  which is invariant under the rotation 
by $\pi/2$ in two dimensions is given as 
\be
 M_R = m_1 \bm{1} + m_2 \Gamma_3 
       + m_3 (\tilde{\gamma}_1 + \tilde{\gamma}_2) 
       + m_4 \Gamma_3(\tilde{\gamma}_1 + \tilde{\gamma}_2), 
\label{total-mass}
\ee
\nid
where $m_1$, $m_2$, $m_3$ and $m_4$ are parameters of each operator 
in Eq.~($\ref{ops}$).  $M_R$ has four eigenvalues  
\ba
 m_1-m_2-\sqrt{2}m_3+\sqrt{2}m_4 , & & 
 m_1-m_2+\sqrt{2}m_3-\sqrt{2}m_4 , \no 
 m_1+m_2-\sqrt{2}m_3-\sqrt{2}m_4 , & & 
 m_1+m_2+\sqrt{2}m_3+\sqrt{2}m_4 .  
\label{eigenvalues}
\ea
\nid
A  4-component spinor should be separated into two 
2-component spinors since a two-dimensional Dirac spinor is composed 
of a 2-component mode  and we keep the rotational invariance 
even under  a finite lattice constant%
\footnote{If one permits  the rotational invariance only after taking 
the continuum limit,  it is not necessary for degeneracy of a heavy
mode  and there are six more cases derived from Eq.~(\ref{eigenvalues}).}. 
Actually all possibilities of this separation are three cases 
and are listed in Table~\ref{3C}.  

\begin{table}[h]
\begin{center}
\renewcommand{\arraystretch}{1.25}
 \begin{tabular}{|c|c|c|c|} \hline 
   & parameter conditions & rotationally invariant mass term & 
  mass eigenvalues \\ \hline 
  case~ 1 & $m_2 = m_3 = 0$ & 
  $M_{R1}=m_1\bm{1}+m_4\Gamma_3(\tilde{\gamma}_1+\tilde{\gamma}_2)$ & 
  $m_1 \pm \sqrt{2}m_4$ \\ \hline 
  case~ 2 & $m_2 = m_4 = 0$ & 
  $M_{R2}=m_1\bm{1}+m_3(\tilde{\gamma}_1+\tilde{\gamma}_2)$ & 
  $m_1 \pm \sqrt{2}m_3$ \\ \hline 
  case~ 3 & $m_3 = m_4 = 0$ & $M_{R3}=m_1\bm{1}+m_2\Gamma_3$ & 
  $m_1 \pm m_2$ \\ \hline 
\end{tabular}
\caption{Three cases for the mass splitting into two spinors.} 
\label{3C}
\end{center}
\end{table}
After the mass splitting,  
we can find  the character of a Dirac spinor under the rotation, 
\be
 \psi(x) \rightarrow  Q\psi(R(x)) , 
\ee
\nid
where 
$Q = e^{(i\pi/4)\sigma_3}
   = \pmatrix{ e^{i\pi/4} & 0 \cr 0 & e^{-i\pi/4}}$.  
Actually in  cases 1 and 2  we can  keep 
the property of a Dirac spinor on lattice.  
By contrast, $\Psi(N)$ acts as a vector not as a spinor 
in case 3.  The properties  of $2$-component spinors 
under the rotation  are summarized in Table \ref{DS}
\footnote{$M_R$ and $V_{12}$ can be diagonalized simultaneously 
because $[M_R,V_{12}] = 0$.}.  

\begin{table}[ht]
\begin{center}
\begin{tabular}{|c|c|c|}
 \hline & $V_{12}^{diag}$ & phase factor of $V_{12}$ \\ \hline
  $\begin{array}{c} ~ \\[-5pt] {\rm case~1} \\[-5pt] ~ \end{array}$ & 
  $\left(\begin{array}{cc} Q & 0 \\ 0 & e^{i\pi}Q^\dag 
   \end{array}\right)$ & 
  $e^{i\vartheta}=e^{i\pi/2}=i$ \\ \hline 
  $\begin{array}{c} ~ \\[-5pt] {\rm case~2} \\[-5pt] ~ \end{array}$ & 
  $\left(\begin{array}{cc} Q & 0 \\ 0 & e^{i\pi}Q^\dag 
   \end{array}\right)$ & 
  $e^{i\vartheta}=e^{i\pi}=-1$ \\ \hline 
  $\begin{array}{c} ~ \\[-5pt] {\rm case~3} \\[-5pt] ~ \end{array}$ & 
  $\left(\begin{array}{cc} Q^2 & 0 \\ 0 & e^{i\pi/2}(Q^\dag)^2 
   \end{array}\right)$ & 
  $e^{i\vartheta}=e^{-i\pi/4}=(1-i)/\sqrt{2}$ 
 \\ \hline 
\end{tabular}
\caption{The properties of  Dirac spinors under the rotation.}
\label{DS}
\end{center}
\end{table}

\section{Pole Analysis and 2-point Functions}    
Our adding terms do not commute with the staggered Dirac operator.  
As a result,  our analysis in the  previous section 
is insufficient to split masses.  We must proceed in 
the pole analysis of  the theory because a pole mass is physical.  
The staggered Dirac operator in the momentum space is written as 
\be
 D_{st}(p)=\sum_\mu\left\{i\gamma_\mu\sin 
 p_\mu+i\tilde{\gamma}_\mu(1-\cos p_\mu)\right\}. 
\ee

Our steps  to find a pole mass are as follows:
(i) set $p_1=0$ and $p_2= i \kappa~({\rm pure ~imaginary})$ of 
the inverse propagator  $D^{-1}$ in the momentum representation 
where our rotationally invariant operators
are included; (ii)  calculate  four eigenvalues $\lambda$ of 
$D^{-1}$; (iii) find values of $\kappa$ in setting $\lambda=0$.  Four 
values of $\kappa$  equal to pole masses.  As mentioned in 
sections 2 and 3, we keep the rotational invariance in our action and 
generate two Dirac spinors with different masses.  
We define $m_1$, $m_2^{\prime}\equiv -i m_2$, 
$m_3^{\prime}\equiv -i m_3$ and $m_4$ as real parameters 
to obtain real pole masses and then denote by 
$m_l$ and $m_h$ the light and heavy Dirac masses, respectively.  
For each three cases results in brief of the pole analysis are 
as follows.  

\paragraph{$\bullet$ case~1} ~ 

\nid
The pole mass is still splitting under $|m_4|<1$.  
It is also found that  we can take a limit 
$|m_h| \rightarrow \infty$ for arbitrary 
$m_l$ by performing $\epsilon \rightarrow 0$ in 
an expression $m_4^2=1-\epsilon\;(0<\epsilon\ll1)$.  

\paragraph{$\bullet$ case~2} ~ 

\nid
The pole mass remains degenerate because the improved term 
$m_3^\prime (\tilde{\gamma}_1 + \tilde{\gamma}_2)$ is absorbed into 
the kinetic term. 

\paragraph{$\bullet$ case~3} ~ 

\nid
This  case  allows  pole masses  to split 
although the rotational property of the eigenmode 
is not a spinor from the discussion of the previous section.

~ \\

Note that it is possible to take the light mass $m_l$ to zero  by 
tuning $m_1$ and $m_4$ only in case 1.  
Solutions of the equation for the pole mass
under the massless condition $m_1^2=2m_4^2$ are determined as 
\be
 \sinh^2 \frac{m_l}{2}=0~,~~ 
 \sinh^2 \displaystyle\frac{m_h}{2} = \frac{2m_4^2}{1-m_4^2} .  
\label{massless-infty-condition}
\ee
In addition, to decouple the heavy mode, we can throw the mass  
up to  infinity.  
Actually from Eq.~(\ref{massless-infty-condition}), 
we  can realize massless and infinity modes as 
Table~\ref{limitmode}  simultaneously.  
Although the formal 
$\Gamma_5$ 
chiral projection which means even-site and odd-site separation of  
fermion modes  is not consistent with the rotational invariance of 
a staggered Dirac action, 
it is found that 
infinity modes can be  separately  put  on even or odd sites
\footnote{The eigenmode of the Dirac operator 
around a massless pole  is not orthogonal to that 
around a heavy mass pole because their Dirac operators are 
different from each other.}.
\begin{table}[h]
\begin{center}
\begin{tabular}{|c|c|c|} 
 \hline & massless modes & infinity modes \\ \hline 
   $\begin{array}{c} ~ \\[4mm] m_4 > 0 \\[4mm] ~  \end{array}$ 
 & $\left(\begin{array}{c} 
     1+\sqrt{2} \\ -1-\sqrt{2} \\ 1 \\ 1 
    \end{array}\right) , \quad 
    \left(\begin{array}{c} 
     1-\sqrt{2} \\ 1-\sqrt{2} \\ -1 \\ 1 
    \end{array}\right)$ 
 & $\left(\begin{array}{c} 1 \\ 0 \\ 0 \\ 0 \end{array}\right) , \quad 
    \left(\begin{array}{c} 0 \\ 0 \\ 0 \\ 1 \end{array}\right)$ 
 \\ \hline 
   $\begin{array}{c} ~ \\[4mm] m_4 < 0 \\[4mm] ~  \end{array}$ 
 & $\left(\begin{array}{c} 
     1-\sqrt{2} \\ -1+\sqrt{2} \\ 1 \\ 1 
    \end{array}\right) , \quad 
    \left(\begin{array}{c} 
     1+\sqrt{2} \\ 1+\sqrt{2} \\ -1 \\ 1 
    \end{array}\right)$
 & $\left(\begin{array}{c} 0 \\ 0 \\ 1 \\ 0 \end{array}\right) , \quad 
    \left(\begin{array}{c} 0 \\ 1 \\ 0 \\ 0 \end{array}\right) $ 
 \\ \hline 
\end{tabular}
\caption{Eigenvectors of the improved Dirac operator in case 1 with  
$m_1^2 = 2 m_4^2$.}  
\label{limitmode}
\end{center}
\end{table}

\section{Summary and Discussion}
We have studied  the mass splitting of two-dimensional 
staggered fermions based on the $SO(4)$ Clifford algebra.  Introducing 
four rotationally invariant operators,
we have analyzed three types of improved staggered Dirac operators 
and found one possibility (case 1) for taking a single mode 
in a two-dimensional free theory.  The case keeps the splitting 
not only in the analysis of the mass matrix itself 
but also in the pole analysis including the kinetic term.  According
to the improvement with respect to the rotational invariance, the  
derived $2$-component modes can be regarded as the ordinary spinor  
under the rotation by $\pi/2$.  Furthermore, one can find 
a massless mode in the case  unexpectedly.  
Our future tasks are  analyses of interacting theories  and 
the extension of our approach to  four dimensions.  
It is crucial that the stability for the massless condition under 
quantum corrections by gauge interactions.  



\end{document}